\documentclass[aps,prd]{revtex4}
\bibliography{apsrev}
\begin{document}

\title{Conformally Invariant Brane-universe 
	       and the Cosmological Constant}

\author{E.I.Guendelman}
\email{guendel@bgumail.bgu.ac.il }
\affiliation{Physics Department, Ben Gurion University,\\
 Beer Sheva, Israel}

\author{E. Spallucci}
\email{spallucci@trieste.infn.it}
\affiliation{Dipartimento di Fisica Teorica, Universit\`a di Trieste,
		          and INFN, Sezione di Trieste}

\begin{abstract}
        A relativistic $3$-brane can be given a conformally invariant,
        gauge-type, formulation provided the embedding space is
	six-dimensional. The implementation of conformal invariance requires the
	use of a modified measure, independent of the metric in the action.
	A brane-world scenario without the need of a cosmological constant in
	$6D$ can be constructed. Thus, no ``old'' cosmological constant problem 
	appears at this level.
\end{abstract}
	\maketitle
	Extended objects of various dimensions are present in the modern
	formulation of string theory. Among the various kind of branes
	a unique role is played by $D$--branes 
	as they can trap the end-points of open strings. $D$-brane inspired 
	cosmological models, commonly termed ``brane--universe'' models, are 
	currently under investigation as they seem to offer a possible solution
	to the longstanding hierarchy problem in gauge theories.\\
	The ``gauge'' formulation of $p$--branes, proposed some years ago
	as an alternative to the standard description
	of relativistic extended objects \cite{aae},\cite{gtm}, is well suited
	to describe this new type of cosmological scenario. Furthermore,
	the description of $p$--branes in terms of associated gauge potentials,
	$W^{\mu_1\dots \mu_{p+1}}$ and $B_{\mu_1\dots \mu_p}$, offers a 
	vantage point to study some specific problem 
	as the one concerning the fine tuning of the cosmological constant.\\
	$3$-branes considered in an embedding $6D$
	space the gauge theory formulation of $3$-branes allows a conformally
	invariant realization \cite{noi}: 
	
	\begin{eqnarray}
	S = &&-\frac{1}{16\pi G_{(5+1)}}\int d^{5+1}x \, \Phi\, g^{{}^{AB}}\,
	R_{{}_{AB}}\left(\, \Gamma\,\right)
	+e^2\,\int d^{5+1}x \,\Phi\,
	\sqrt{-\frac{1}{2\times 4!}g_{{}_{AE}}\dots g_{{}_{DH}}\, W^{{}^{ABCD}}
	\,
	W^{{}^{EFGH}}\,}+\nonumber\\
	&&-\frac{1}{4!}\,\int d^{5+1}x \,\sqrt{-g_{(5+1)}}\, W^{{}^{EFGH}}
	\partial_{[\, {}_{E}}\, B_{{}_{FGH}\,]}\label{confact}\\
	&&\Phi\equiv \epsilon^{A_1\dots A_6}\epsilon_{a_1\dots a_6}
	\partial_{A_1}\phi^{a_1}\dots \partial_{A_6}\phi^{a_6}
	\end{eqnarray}

	Essential elements necessary to implement
	conformal invariance are the introduction of a measure of integration 
	$\Phi$ in the action which is independent of the metric
	\cite{g3},\cite{g4},\cite{g5}, \cite{g6} and a first order formulation
	of the gravitational action.
	Brane world scenarios in general are concerned with the possibility that
	our universe is built out of one or more $3$-branes living in some
	higher dimensional space, plus some bulk component \cite{7},
	\cite{8}, \cite{9}, \cite{10}, \cite{11}.
	In particular, $3$-branes embedded in $6D$ space 
        \cite{12},  \cite{13}, \cite{14}, \cite{15}, \cite{16}.
	induce curvature only in the extra dimensions. However, although the
	branes themselves do not curve the observed four dimensions, the bulk
	components of matter do, and they have to be \textit{fine tuned} in order to get
	(almost)zero four dimensional vacuum energy.\\
	In order to solve this problem we propose to
	incorporate the ``brane-like features'', that are quite good, in what
	concerns the cosmological constant problem into the ``bulk'' part of the
	brane scenario as well. In this way both bulk and singular brane
	contributions will share the fundamental feature of curving only the
	extra dimensions. The gauge field formulation of $3$-branes in $6D$ 
        is ideally suited for such a program.
	The action (\ref{confact}) leads to a set of brane-world solutions, 
	including the ones presented in \cite{17} as very particular 
	cases,
	\begin{eqnarray}
	&& ds^2=\eta_{\mu\nu}\, dx^\mu_{||}\, dx^\nu_{||}+ 
	\psi\left(\, r \right)\left[\, dr^2 + r^2\, d\phi^2\,\right]
	\label{lelem}\\
	&& \psi=\frac{4\alpha^2 b^2}{r^2}\left[\, \left(\, \frac{r}{r_0}
	\right)^\alpha +  \left(\, \frac{r}{r_0} \right)^{-\alpha}\,\right]^{-2}
	\\
	&&\alpha \equiv 1-4 G_{(5+1)}T\ ,\qquad
	b^2\equiv \frac{\sqrt 2}{16\pi G_{(5+1)}B_0}
	\end{eqnarray}
	 Notice that the $4D$ part does not suffer curvature.$B_0$ and $T$ are 
	 constants.\\
	In this communication we have discussed how the gauge formulation of 
	branes can be used in the framework of ``brane world'' scenarios.\\
	The formulation of $3$-branes in a six-dimensional target spacetime
	can be made in a conformally invariant way. This is possible for
	extended objects in case the target spacetime has two more dimensions
	than the extended object itself. \\
	This conformal invariance is intimately related to fact that the branes 
	( or equivalently the associated gauge fields ) only curve the manifold
	orthogonal to the brane, the extra-dimensions. No fine tuning of a
	$6D$ cosmological constant is needed in this case. Therefore, no ``old
	cosmological constant problem'' , as Weinberg has defined it \cite{sw},
	appears.\\ While a conformally invariant
	formulation of the brane alone was already achieved in \cite{weyldp}, 
	it is only after the inclusion of gravity into the conformal invariance
	that the formulation can have an impact into the question of the 
	cosmological constant problem. In our model we are able to formulate
	the brane plus gravity in a conformally invariant fashion provided
	the $3$-brane is embedded in a six dimensional space.

\end{document}